\documentclass[a4paper, 11pt]{amsart}
\usepackage{amsmath,amssymb}
\begin{document}
\title[Super-Heisenberg group and harmonic superanalysis]{$Q$-representations and unitary representations of the super-Heisenberg group and harmonic superanalysis}
\maketitle
\author{\centerline{\sc Andrzej M.Frydryszak}}

\address{Institute of Theoretical Physics, University of Wroclaw, Wroclaw, Poland}


\begin{abstract}
We juxtapose two approaches to the representations of the super-Heisenberg group. Physical one,  sometimes called concrete approach, based on the super-wave functions depending on the anti-commuting variables, yielding the harmonic superanalysis and recently developed strict theory of unitary representations of the nilpotent super Lie groups covering the unitary representations of the super-Heisenberg group.
\end{abstract}

\section{Introduction}
The Heisenberg group plays distinguished role as in theoretical physics as in mathematics, being a fundamental
building block of many theories and formalisms \cite{folland, howe}. From the physical point of view it comes from the linearization of the group of canonical transformations. Its representations, in fact, arise from the quantization procedure - the canonical quantization. This procedure has its own difficulties, but
al the l quantum physics we know and which is supported by experiment, was discovered by the use of this tool, despite the fact that recent decades brought vivid development of new approaches, to mention the deformation quantization only.
There are lots of generalizations and applications of the original Heisenberg group  and of the harmonic analysis on it, depending on the mathematical or physical context (discrete, supersymmetric, deformed). Here we want to focus only on the generalization coming from the supersymmetry in the context of physics and related to it supermathematics \cite{manin, leites}. Firstly we shall discuss supermatrix version of it, coming from the supersymmetric mechanics and in particular from the supersymmetric model of the  harmonic oscillator. Such a concrete approach is good enough to produce  superized version of the harmonic analysis on the super-Heisenberg group, special superfunctions (super-Hermite polynomials, super-Laguerre etc.). It is related to the so called super-Schroedinger quantization. Within supersymmetric approach it is possible to introduce two types of super-Heisenberg group namely the even super-Heisenberg group and the odd super-Heisenberg group. The even super-Heisenberg group is the one usually discussed in the literature, the odd one, is less frequent. To write  (super)representations of the odd super-Heisenberg group one has to introduce a new algebra of "numbers" with the odd unit and and odd imaginary unit in the real and the complex case respectively.
Such structure, named the algebra of oddons,  has been defined in Ref. \cite{af-oh}. Its particular version is also known under the name super skew field \cite{kwok} and plays important role in algebraic supergeometry \cite{kwok}.

The plan of the paper is the following. Firstly we discuss the super-matrix formulation of the even and odd super Heisenberg group and its $Q$-representations resulting from the super-Schr\"{o}dinger quantization, then we briefly comment recent developments in the strict theory of unitary representations of the super Heisenberg group defined as a super Harish-Chandra pair.
%
\section{Super-Heisenberg group}
In the so called concrete approach the super-Heisenberg group is introduced as a supermatrix group,
 related to supersymmetric mechanics formalism in analogy to the way that conventional Heisenberg
group arrises in the context of the phase space formulation of classical mechanics \cite{folland}.
Let us briefly recall this way of defining Heisenberg group and some points of the Schr\"{o}dinger
quantization procedure which yields the harmonic analysis on Heisenberg group. The effect of the
canonical quantization procedure is summarized by the Stone-von Neumann theorem.

For the super-Heisenberg group one can follow similar lines, in particular, so called Grassmannian Fourier transform can be defined and superized versions of the Hermite and Laguerre polynomials obtained. The super-Hermite polynomials arise in the super-Sr\"{o}dinger quantization of the  fermionic super-oscillator \cite{af-osc}. To fix the notation let us remind some fact concerning the Heisenberg group and phase space, this will form a background to the generalization to the super-Heisenberg group case.
%
%
\subsection{Heisenberg group}

Let us consider conventional phase space $\mathcal{P}_{2n}\ni(p,q)$ with  a
symplectic form $\Omega$: $\Omega((p,q),(p',q'))=pq'-p'q$ which relates Poisson bracket of classical phase space observables with Hamiltonian vector fields
\begin{equation}
\{f,g\}=\Omega(X_f,X_g)
\end{equation}
and moreover
$$
[X_f, X_g]=X_{\{f,g\}},\quad \{p^j,\,q_i\}=\delta^j_i
$$

In the extended phase space $\mathcal{P}_{2n+1}\ni(p,q,t)$ one can define  Heisenberg algebra by commutation relations
\begin{equation}
 [(p,q,t),(p',q',t')]=(0,0,pq'-p'q)=(0,0,\Omega((p,q),(p',q'))),
\end{equation}
therefore
 \begin{equation}
 [P_j, Q_k]=\delta_{jk}T.
 \end{equation}
Such commutation relations can be explicitly realized \cite{folland} by matrices
\begin{align}
m(p,q,t)=
 \begin{pmatrix}
 0 & p & t \\
 0 & 0 & q \\
 0 & 0 & 0
 \end{pmatrix}
 =m(p,q,t)m(p',q',t')=m(0, 0, pq'), 
\end{align}
There are two isomorphic forms of the group realization. The first  one  given by
\begin{align}
M(p,q,t) =1+m(p,q,t)
\end{align}
and yields the following form of group multiplication
\begin{equation}
 M(p,q,t)M(p',q',t')=M(p+p',q+q',t+t'+pq').
 \end{equation}
 The second one is obtained from the exponentiation
 \begin{equation}
 e^{m(p,q,t)}=1+m(p,q,t)+\frac{1}{2}pq=M(p,q,t+\frac{1}{2}pq)
 \end{equation}
 and produces more symmetric form of the multiplication
 \begin{equation}
 e^{m(p,q,t)}e^{m(p',q',t')}=e^{m(p+p',q+q',t+t'+\frac{1}{2}(pq'-p'q))}
 \end{equation}
 The group law can be written in a compact form showing the relation to the symplectic geometry of the phase space, namely
 \begin{equation}
 (p,q,t)\circ(p',q',t')=\left( p+p',q+q',t+t'+\frac{1}{2}\Omega((p,q),(p',q'))\right)
 \end{equation}
 
The Schr\"{o}dinger quantization procedure uses the following construction. We take as the quantum state space of the system the Hilbert space  $\mathcal{H}=L^2(R^n, d\mu)$, where the $R^n$ is the configuration space of a classical system, and realize the canonical commutation relations: $[\hat{X},\, \hat{P}]=i\hbar$:
\begin{equation}
\hat{X}: f(x)\mapsto x f(x),
\quad\quad
\hat{P}: f(x)\mapsto -i\hbar \frac{\partial f}{\partial x} (x)
\end{equation}
It turns out that his realization is canonical in generic sense. Namely, we have that
\begin{eqnarray}
\hat{X}\rightsquigarrow e^{iq\hat{X}}, &\quad\quad& e^{iq\hat{X}}: f(x)\mapsto e^{iqx} f(x)\\
\hat{P}\rightsquigarrow e^{ip\hat{P}}, &\quad\quad& e^{ip\hat{P}}: f(x)\mapsto  f(x+\hbar p),
\end{eqnarray}
and for elements of the Heisenberg group
\begin{align}\nonumber
H_n\ni M(p,q,t) =
 \begin{pmatrix}
 1 & p & t \\
 0 & 1 & q \\
 0 & 0 & 1
 \end{pmatrix}
\end{align}
one obtains representation
 \begin{equation}
 \rho: H_n\rightarrow U{\mathcal{H}} \quad\quad M(0,q,0) \mapsto e^{iqx}  \quad\quad M(0,0,p) \mapsto e^{ip\hat{P}}
 \end{equation}
with
\begin{equation}
\rho_h(p,q,t)=e^{2\pi iht}e^{2\pi i(h p D+qX)}.
\end{equation}
The $ U{\mathcal{H}}$ denotes the set of unitary mappings on the $\mathcal{H}$.
The $\rho_h(p,q,t)$, $h\in R$, $h\neq 0$ forms one parameter family of the irreducible unitary representations of the $H_n$. As it was already mentioned this construction is canonical in the sense given by the Stone-von Neumann theorem: {\em
Any irreducible unitary representation of $H_n$ that is nontrivial on the center is equivalent to some $\rho_h$.}
%
%
\subsection{The even super-Heisenberg group}
To introduce the super-Heisenberg group in the supermatrix form let us follow  lines, similar to the ones sketched above. Namely, for supersymmetric system one introduces the graded phase space $\mathcal{P}_{(2n|2m)}\ni (p,q;\Pi, \Theta)$ with  $\Pi=(\Pi^j)$, $\Theta=(\Theta_j)$, where $j=1,\dots, m$ ; with the parity $|\Theta_j|=1=|\Pi^j|$. For physical models the $\mathcal{P}_{(2n|2m)}$ is $Z_2$-graded module over $Z_2$ graded algebra $Q$ \cite{af-lj}. We introduce the supersymplectic form  $B$
 \begin{equation}
 B(v,v')=-(-1)^{|v||v'|}B(v',v), \mbox{for homogenous } \quad v, v' 
 \end{equation}
 The even part of  the $\mathcal{P}_{(2n|2m)}$ contains usual phase space, therefore let us here restrict to the new odd sector $(\Pi, \Theta)\in\mathcal{P}_{2m}$. In the odd basis $\{ E_j \}_1^{2m}$ one can write
\begin{equation}
v=\Pi^j E_j+\Theta_j E^{m+j}
\end{equation}
and
\begin{equation}
B(v,\, v')=\Pi^j\Theta'_j+\Theta_j\Pi'^j
\end{equation}
The generalization of the Heisenberg group related to this sector we shall call the fermionic Heisenberg group. It is a part of the whole super-Heisenberg group.

Let us consider the super-Poison bracket:  $f,g,h\in\mathcal{F}(\mathcal{P}_{2m})$, where $f(\Pi, \Theta)$,
\begin{equation}
\{f,\,g\}=-(-1)^{|f||g|}\{g,\,f\}
\end{equation}
\begin{equation}
(-1)^{|r||h|}\{f\, ,\{g,\,h\}\}+(-1)^{|g||f|}\{g\, ,\{h,\,f\}\}+(-1)^{|h||g|}\{h\, ,\{f,\,g\}\}=0
\end{equation}
for canonical variables it gives
\begin{equation}
\{\Pi^j\, ,\Theta_i\}=\delta^j_i.
\end{equation}
As before we shall pass to the extended phase space $(\Pi,\, \Theta,\, t)\in\mathcal{P}_{(1|2m)}$
and generalized relations
\begin{equation}
[(\Pi,\,\Theta, t),\, (\Pi',\,\Theta', t')]_+=(0,\, 0, \, \Pi\Theta'+\Theta\Pi')
\end{equation}
yielding the supercommutator
\begin{equation}
[\hat{\Pi},\, \hat{\Theta}]_+=\hat{T}.
\end{equation}
Natural supermatrix realization of fermionic-Heisenberg algebra has the following form
\begin{align}
\mu(\Pi,\Theta,t) \longrightarrow
\left(
\begin{array}{cc|c}
0 & t &\Pi \\
0 & 0 & 0 \\
 \hline
0 & \Theta & 0
\end{array}
\right)
; \quad\quad \mu(\Pi,\Theta,t)^3 =0
\end{align}
One of the possible forms of the supermatrix realization of the fermionic Heisenberg group is given by
\begin{align}
M(\Pi,\Theta,t)=1+\mu(\Pi,\Theta,t)
\end{align}
and it yields the "polarized form" of the multiplication:
$$
M(\Pi,\Theta,t)\cdot M(\Pi',\Theta',t')=M(\Pi+\Pi',\Theta+\Theta', t+t'+\Pi\Theta')
$$
The supermatrix realization of the whole super-Heisenberg  group $sH$ with polarized form of the  multiplication reads now
\begin{align}
(p,q,\Pi,\Theta,t) =
\left(
\begin{array}{ccc|c}
1 & p & t &\Pi \\
0 & 1 & q & 0 \\
0 & 0 & 1 & 0 \\
\hline
0 & 0 & \Theta & 1
\end{array}
\right)
\end{align}
As before one can take realization coming from the exponentiation, which produces "unpolarized" form of multiplication and relates it to the super-symplectic geometry of the phase super-space
\begin{eqnarray}
&&(p,q,\Pi,\Theta,t)\diamond (p',q',\Pi',\Theta',t')= \\\nonumber
&&(p+p',q+q',\Pi+\Pi',\Theta+\Theta', t+t'+\frac{1}{2}(pq'-qp')+\frac{1}{2}(\Pi\Theta'+\Theta\Pi')).
\end{eqnarray}

\section{Super-Schr{\"o}dinger representation and harmonic superanalysis}
The super-Schr\"{o}dinger approach to quantization of a supersymmetric system is the  generalization
of the conventional method of quantization to the situation where one has to deal with anticommuting
co-ordinates. This kind of reasoning is widely used in (relativistic) supersymmetric field theory within the superfield formalism.
Here, for supermechanics, one considers a generalization  of the Hilbert space of the square integrable functions defined on the configuration space of the classical system,  what gives the superspace of wave superfunctions integrated in the sense of the Berezin integral.  One of possible structures realizing this goal is the $Z_2$-graded bimodule over $Z_2$-graded Banach-Grassmann algebra $Q$\cite{af-lj}. This kind of an approach is a natural one and in fact was the first considered in the literature by Valle \cite{valle} and falls into the scheme developed by  DeWitt \cite{dewitt}. Here we shall use a slight mutation of the super Hilbert space considered by DeWitt \cite{dewitt}, it was  introduced in Ref. \cite{af-lj} within  the super-GNS construction. This super Hilbert space, as it was mentioned above, is the $Z_2$-graded $Q$ bimodule, with generalized scalar product with values in $Q$-algebra \cite{af-lj}.

To fix the notation, let
\begin{equation}
q\in Q=Q_0\oplus Q_1; \quad\quad q_sq'_r=(-1)^{sr}q'_rq_s, \quad q_s\in Q_s
\end{equation}
with the following extension of complex conjugation \cite{af-lj}
\begin{equation}\label{conjugation}
q^*=\bar{q}_0-i\bar{q}_1; \quad\quad (qq')^*=q'^*q^*
\end{equation}
Functions $\mathcal{F}(\zeta)$, $\zeta=(\zeta_1, \dots, \zeta_m)$, $|\zeta_j|=1$; $d\zeta=d\zeta_m\dots d\zeta_1$, where $\zeta_i \in Q_1$
$$
<f,g>=\int d\zeta f^*(\zeta)g(\zeta)
$$
and $f(\zeta)=\sum_{k,I_k}f_{I_k}\zeta^{I_k}$, where $I_k$ denotes strictly ordered multi-index. Operators conjugated, with respect the above $Q$-scalar product, to the derivation with respect to $\zeta$ and multiplication by $\zeta$ are the following
$$
\partial_{\zeta}^+=i\partial_{\zeta}, \quad\quad \hat{\zeta}^+=-i\zeta
$$

Let $D_j=-i\partial_{\zeta_j}$  and $X^j=\zeta^j$, the operator
$\Pi^j D_j+\Theta_j X^j$ is formally  selfadjoint in $(\mathcal{F}(\zeta), \, <\, ,\,>)$ and
\begin{equation}
\exp{[i(t+\Theta X+\Pi D)]} f(\zeta)=\exp{[i(t+\Theta\zeta+\frac{1}{2}\Theta\Pi)]} f(\zeta+\Pi)
\end{equation}
or equivalently
\begin{equation}
e^{i(\Theta X+\Pi D)}=e^{\frac{1}{2}\Theta\Pi} e^{i\Theta X} e^{i\Pi D}.
\end{equation}
The representation  $\pi_1$ of the fermionic Heisenberg group $FH_m$  defined as
$$
\pi_1 (\Pi,\, \Theta, \,t) f(\zeta)=e^{i(t+\Theta \zeta +\frac{1}{2}\Theta\Pi)}f(\zeta+\Pi)
$$
is $Q$-representation of $FH_m$. Such  $Q$-representation is defined as a generalization of the notion of representation to the $\star-Q$-algebra case, cf. Ref.  \cite{af-lj}.
%
\subsection{Grassmannian Fourier-Wigner transform}
To illustrate how the generalization of the harmonic analysis on the Heisenberg group can be formulated for the super-Heisenberg group let us restrict to the fermionic Heisenberg group. In this case one can introduce the so called  Grassmannian Fourier-Wigner transform \cite{af-fh}. Taking the
 supermatrix elements of the representation $\pi_1$  we get
\begin{equation}
\mathcal{M}(\Pi,\, \Theta)=<f, \pi_1(\Pi,\, \Theta)g>, \quad\quad f,g\in\mathcal{F}(\zeta),
\end{equation}
 Let $V(f,g)$ be a function defined as
\begin{eqnarray}\nonumber
V(f,g)(\Pi,\,\Theta)&=&\mathcal{M}(\Pi,\, \Theta)=<f,\,e^{i(\Theta X+\Pi D)}g>=\\ \nonumber
&=&\int d\zeta f^*(\zeta-\frac{1}{2}\Pi)e^{i\Theta\zeta}g(\zeta+\frac{1}{2}\Pi)
\end{eqnarray}
In analogy to the conventional case \cite{folland} above relation gives rise to the definition of
the so called Grassmannian Fourier-Wigner transform (GFW)  \cite{af-fh} or in other words the super Fourier-Wigner transform, when considered for the whole super-Heisenberg group. Namely,
\begin{equation}
V:\mathcal{F}_m(\zeta)\times \mathcal{F}_m(\zeta)\rightarrow \mathcal{F}_{2m}(\zeta)
\end{equation}
The super Fourier-Wigner transform has analogous properties to the original one, of course, modified by the presence of the $Z_2$-gradation. For example, if  $f_1, \,f_2, \,g_1, \,g_2\in\mathcal{F}_m(\zeta)$  we have
\begin{equation}
<V(f_1,\, g_1),\, V(f_2,\, g_2)>=(-1)^{|f_1||f_2|+(|f_1|+|f_2|)|g_2|}<g_1,\, g_2><f_1,\, f_2>^*.
\end{equation}
Moreover,
let $\alpha,\, \beta,\, \gamma,\, \delta\in Q_1$, then
\begin{eqnarray}
&&V(\pi_1(\gamma, \delta)f, \pi_1(\alpha, \beta)g)(\Pi, \Theta)=\\\nonumber
&&\exp{[-\frac{1}{2}(-\Pi\beta -\Theta\alpha+\gamma\Pi+\delta\Theta+\gamma\beta+\delta\alpha)]}\\\nonumber
&&V(f, g)(\Pi+\alpha-\gamma, \Theta+\beta-\delta).
\end{eqnarray}
%
%
\subsection{Grassmannian Bargmann transform}
The Grassmannian Bargmann transform is the specific GFW-transform, defined for spaces with the even dimension $m$. In such a case one can consider the Grassmannian Gausian $\omega_0\in\mathcal{F}_m(\zeta)$; $m=2k$
\begin{equation}
\omega_0=(Pf G)^{-\frac{1}{2}} e^{\frac{1}{2}\zeta G\zeta}, \quad\quad G_{ij}=-G_{ji},
\end{equation}
where $Pf G$ denotes the Pfaffian of the antisymmetrix matrix $G$. The $\omega_0$ is normalized with respect to the $Q$-scalar product
$$
<\omega_0,\, \omega_0>=1.
$$
Fixing  $\omega_0$ as one of the entries of the super Fourier-Wigner transform we obtain
\begin{equation}
V(\omega_0, f)(\Pi, \Theta)=(Pf G)^{-\frac{1}{2}} \int d\zeta e^{(\zeta-\Pi)G(\zeta-\Pi)}e^{i\Theta\zeta-\frac{i}{2}\Theta\Pi}f(\zeta)
\end{equation}
In terms of the new variable $z_k=G_{kj}\Pi^j+i\Theta_k$ we have
\begin{equation}
V(\omega_0, f)(\Pi,\Theta)=(Pf G)^{-\frac{1}{2}} e^{-\frac{1}{4}z^*G^{-1}z}\int d \zeta e^{\frac{1}{2} \zeta G\zeta-\zeta z-\frac{1}{4}z^*G^{-1}z}f(\zeta)
\end{equation}
This formula gives rise to the definition of the Grassmannian Bargmann (super-Bargmann) transform \cite{af-fh}
\begin{equation}
({\bf B}f)(z)\equiv 2^{-\frac{n}{2}}\int d\zeta e^{\frac{1}{2} \zeta G\zeta-\zeta z-\frac{1}{4}z^*G^{-1}z}f(\zeta)
\end{equation}
Observing that  $Im(-\frac{i}{2}z^*G^{-1}z')=B(v,v')$ and using identification  $z=G\Pi+i\Theta \rightarrow (\Pi, \Theta)$ one can rewrite super-Heisenberg group law in the form
 $(z,t)\cdot (z',t')=(z+z', t+t'+\frac{1}{2}Im(-\frac{i}{2}z^*G^{-1}z'))$. Finally, as in the conventional case \cite{folland} the transferred representation $\beta$ can be defined \cite{af-fh} i.e. $\beta(z,t)\circ {\bf B}={\bf B}\circ \pi_1(\Pi,\Theta, t)$.
denoting $||z||^2=\frac{1}{2}z^*G^{-1}z$  we get relation
\begin{equation}
V(\omega_0,f)(\Pi,\Theta)=e^{-\frac{1}{2}||z||^2}({\bf B}f)(z)
\end{equation}
In view of the above relations it is natural to consider the notion of the Grassmannian Bargmann-Fock space (the super Bargmann-Fock space) $\mathcal{F}(z)$ . In the $\mathcal{F}(z)$  we define Q-scalar product
\begin{equation}
<f,g>_{\mathcal{F}}=\int |dz| e^{-\frac{1}{2}||z||^2} f^*(z)g(z), \quad\quad |dz|=-(\frac{i}{2})^n dzdz^*
\end{equation}
In particular monomials in $z$-variables are orthonormal with respect to this $Q$-scalar product i.e. $<z_{I_k},z_{J_l}>_{\mathcal{F}}=\delta_{I_k J_l}$.  For the $Q$-representation $\beta$, using $w=G\rho+i\sigma$  one can show that
\begin{equation}
(\beta(w){\bf B}f)(z)=e^{-\frac{i}{2} ||w||^2}e^{-\frac{i}{2}zG^{-1}w^*}{\bf B}f(z)
\end{equation}
%
%
\subsection{Grassmannian special functions}%
The Grassmannian Hermite (the super Hermite) polynomials arise in the super-Schr\"{o}dinger quantization of the fermionic harmonic oscillator, they can be defined by the Rodrigues-like formula \cite{af-gsf}. Namely, the components of the Grassmannian Hermite polynomials are given by
\begin{equation}
h_k^{I_k}=H_k e^{-\frac{1}{2}\zeta G\zeta} \partial^{I_k} e^{\zeta G\zeta }. 
\end{equation}
It is interesting that the super-Bargmann transform, as in the conventional case,  relates polynomial base in the   $\mathcal{F}(z)$ and the super-Hermite polynomials in the super-Fock $Q$-module.
$({\bf B}h_k)_{I_k}(z)=2^{\frac{n}{2}} H_k z_{I_k}$. Moreover, using the super Fourier-Wigner transform of the super-Hermite polynomials one can obtain super-Laguerre polynomials \cite{af-fh}.

\section{Odd super-Heisenberg group}
It this section we shall discuss the odd super-Heisenberg group. Let us recall that generic super-phase
 space can be of two types, depending wheter the supersymmetric system is defined by the even or the odd Hamiltonian. The models described by the even Hamiltonian are more frequent in the supersymmetry.
 In such models parity of super-coordinates and canonical super-momenta is the same: $|q_i|=|p^i|=0 $ and $|\Theta_j|=|\Pi^j|=1$. This case is covered by the super-symplectic geometry
 However, as it was recognized by Leites \cite{leites-per} there exists another option for nontrivial geometry defined by the odd structure, the so called periplectic form. Systems falling into this type are defined by the odd Hamiltonian and parity of the super-coordinates and canonical super-momenta is opposite \cite{af-odd} i.e.  $|q_i|=0=|\Pi^i|+1$(mod 2) and $ |\Theta_j|=1=|p^j|+1 $. Hence, in generall we can consider the following sectors
\begin{align}
\begin{array}{c|c}
(p,q) & (p,\Theta) \\
&\\[-3mm]
\hline
&\\[-3mm]
(\Pi,q)  & (\Pi,\Theta) \\
\end{array}
\end{align}
The (even) super-Heisenberg group is defined on the "diagonal" of the above diagram, the odd super-Heisenberg group we want to describe now is defined on the "off diagonal" sectors. We wish to fix our attention on the one of them: $(\Theta_j ,p^j)$. Both odd sectors can be mapped one into the other by the parity shift operator, preserving the type of the periplectic  supergeometry they have. Related part of the odd super-Heisenberg group we shall briefly call the odd Heisenberg group $\mathcal{OH}_m$.

To introduce such an odd Heisenberg group we need the odd super-symplectic form  $\stackrel{(1)}{B}(p,\Theta)$,  with property that $\stackrel{(1)}{B}(v_s,w_r)\in Q_{s+r+1}$.
Let us define it as a nondegenerate $Q$-left linear $B$ form, such that
\begin{equation}
\stackrel{(1)}{B}(v_r,w_s)=-(-1)^{(r+1)(s+1)}\stackrel{(1)}{B}(w_s,v_r)
\end{equation}
The  $\stackrel{(1)}{B}$  defined in natural base can be written as
 \begin{equation}
 \stackrel{(1)}{B}(v,v')=p^j \Theta'_j - \Theta_j p'^j
 \end{equation}
For the odd supermechanical systems there is naturally defined the odd Poisson bracket, having the properties of the odd graded bracket (called also antibracket) i.e.
\begin{equation}
A=A_0 \oplus A_1; \quad[\cdot,\, \cdot]_1:A_s\times A_r \rightarrow A_{s+r+1}
\end{equation}
\begin{equation}
[a,\, b]_1=-(-1)^{(a+1)(b+1)}[b,\, a]_1
\end{equation}
\begin{equation}
\sum_{cycl} (-1)^{(a+1)(b+1)} [a,\, [b,\, c]_1]_1=0
\end{equation}
Let us here note that  in the graded algebra with the odd associative multiplication, such an odd graded bracket can be realized as graded odd commutator, what we will need to proceed with the canonical super-quantization procedure for the odd systems.
To consistently extend the odd super-phase space we have to introduce the odd "time" parameter $\tau$ to have the odd Heisenberg group law defined in the connection to the odd geometry
$
(v,\tau)\bullet(v',\tau ')=(v+v',\, \tau+\tau' +\frac{1}{2}\stackrel{(1)}{B}(v,v')).
$
The odd time can be realized as $Q$-number i.e. $\tau \in Q_1$, but in that case $\tau^2=0$ and we lose the physical interpretation and possibility to use $\tau$ as an evolution parameter. On the other hand, as we have already mentioned the odd multiplication seems to be necessary to consistently quantize the odd systems. To solve these difficulties one has to introduce a new structure. In the next section we shall show that the so called algebra of oddons \cite{af-oh} can fulfill our needs.
%

\subsection{Oddons}
The oddons were introduced in Ref. \cite{af-fh} in the context of the harmonic superanalysis on the odd Heisenberg group, but it turns out that similar structure appears in the algebraic supergeometry \cite{kwok}. We shall briefly describe real and complex oddons.
\subsubsection{Real Oddons}
The algebra of real oddons $\mathcal{O_R}$ is generated by the unit $1\in Q_R$ and a new generator  $\hat{1}$ such that:
\begin{equation}
\hat{1}^2=1, \quad |\hat{1}|=1, \mbox{i.e.}\quad  \hat{1}q_s=(-1)^s q_s \hat{1}
\end{equation}
\begin{equation}
\mathcal{O}_R=Q_R \oplus \hat{1}Q_R
\end{equation}
Any oddon can be written as $r=a+\hat{1}b$, $a, b \in{Q_{R}}$ or as a pair $(a,b)\in Q_R^{(1|1)}$ and can be decomposed into the homogenous parts.
\begin{equation}
a+\hat{1}b=(a_0+\hat{1}b_1)+(a_1+\hat{1}b_0), \quad\quad a_s, b_{s}\in Q_s; \quad s=0,1
 \end{equation}
 With the use of parity mapping $J$ on the $Q$ algebra, the product of two real oddons can be written in the following form
\begin{equation}
rr'=(a+\hat{1}b)(a'+\hat{1}b')=aa'+J(b)b'+\hat{1}(J(a)b'+ba'),
\end{equation}
 where $ J(a)=J(a_0+a_1)=a_0-a_1$.
This multiplication is associative, but because of the presence of $\hat{1}$,  $\hat{1}^2\neq 0$  it   is not graded commutative, what can be easily verified. Interesting sub algebras in the oddonic algebra are the following:
\begin{itemize}
\item the subalgebra of even oddons $r_0=a_0+ \hat{1}b_1$; $a_0\in Q_0$, $b_1\in Q_1$
\item the subalgebra oddons over the body of $Q_R$ i.e. over real numbers $r=a+\hat{1}b$;
$a,b\in R$. In this case $r\in R^{(1|1)}$.
\end{itemize}
 One can show, that for homogeneous oddons with nonvanishing body of the even component there exists the inverse.
 Finally, let us observe that in the oddon algebra we can introduce the new odd product
\begin{equation}
r\ast r'=r\cdot\hat{1}\cdot r',
\end{equation}
which we shall use to generalize structures needed for odd super-Heisenberg group and its super-Schr\"{o}dinger quantization.

The reall Oddons in a way, are graded generalization of so called split numbers or double numbers.
One can also consider in this context a counterpart of the graded dual numbers, where additional generator is nilpotent.
\subsubsection{Complex Oddons}
To separate information about new odd unit $\hat{1}$ and the complexification we have firstly described real version of oddons. The definition of the complex oddons requires the introduction of an "odd imaginary unit" $\hat{\iota}$ such that
\begin{equation}
\hat{\iota}^2=-1, \quad |\hat{\iota}|=1,\quad\quad \mbox{i.e.}\quad  \hat{\iota}q_s=(-1)^s q_s \hat{\iota}
\end{equation}
So, this time we have that $\mathcal{O}=Q \oplus \hat{\iota}Q$. As for real oddons, here also the multiplication
\begin{equation}
zz'=(a+\hat{\iota}b)(a'+\hat{\iota}b')=aa'-J(b)b'+\hat{\iota}(J(a)b'+ba')
\end{equation}
is associative, but is not graded commutative. As in the case of real oddons, for homogeneous elements with nonvanishing body of the even component there exists the inverse. There are also distinguished subalgebras similar to the mentioned above subalgebras of the real oddons.
\subsection{Odd Heisenberg group}
Let us extend the phase space of the odd system by the odd time parameter $\tau$, but with $\tau$ taking oddonic values i.e. $\tau=t\hat{1}\in\mathcal{O}$ and $t\in R$. The group law for the odd Heisenberg group now can be defined by
\begin{equation}
(v,\tau)\bullet(v',\tau ')=(v+v',\, \tau+\tau' +\frac{1}{2}\stackrel{(1)}{B}(v,v'))
\end{equation}
The supermatrix realization (with the oddonic entries) of the generator takes similar form to the previous one. We have
\begin{align}
\mu(p,\, \Theta,\, \tau)=
\left(
\begin{array}{cc|c}
0 & p &\tau \\
0 & 0 & \Theta \\
\hline
0 &  0& 0
\end{array}
\right)
\end{align}
With the use of the odd multiplication we can define odd exponents $e_*^q$  and realize the odd Heisenberg group with multiplication defined by the odd supersymplectic form
\begin{equation}
e_*^{\mu(p,\Theta,\tau)}\ast e_*^{\mu(p',\Theta',\tau')}=
e_*^{\mu(p+p',\Theta+\Theta',\tau+\tau'+\frac{1}{2}(p\Theta'-\Theta p'))}
\end{equation}
%
\section{Oddonic super-Schr\"{o}dinger representation}
The $\mathcal{O}$-representations of the odd Heisenberg group have similar structure to the
super-Schr\"{o}dinger representations of the even super-Heisenberg group. Here we have to consider
relevant objects over oddons instead of the $Q$ algebra. In particular
\begin{itemize}
\item $<f,g>=\int d\zeta f^*(\zeta)g(\zeta)$, $\mathcal{O_C}$-valued,
\item $D_j=\hat{\iota}\partial_{\zeta_j}$ and $X^j=\zeta^j$, $\zeta\in Q_1$
\item $p^i=\hat{1}\Pi^i$,  $\Pi^i\in Q_1$
\end{itemize}
The odd Heisenberg group  $\mathcal{O}$-representation $\pi (p, \Theta ,t)$  is based on the relation
\begin{equation}
exp_*{[\hat{\iota}(t+\Theta X+\Pi D)]} f(\zeta)=exp_*{[i(t+\Theta\zeta+\frac{1}{2}\Theta\ast p)]} f(\zeta+\hat{1}p)
\end{equation}
Because  $p^i=\hat{1}\Pi^i$  the algebraic form of  formulas obtained for the odd Heisenberg group
turns out to be  preserved. In particular, the  Grassmannian Fourier-Wigner transform in
the oddonic case takes the following form
\begin{equation}
V(f,g)(p,\,\Theta)=\int d\zeta f^*(\zeta-\frac{1}{2}\hat{1}p)e^{i\Theta\zeta} g(\zeta+\frac{1}{2}\hat{1}p)
\end{equation}
The odd exponent allows to define the odd Gaussian $e_*^{\zeta\hat{G}\zeta}$, where $\hat{G}=\hat{1}G$, $G_{ij}=-G_{ji}$ and to introduce the odd Bargmann transform
\begin{equation}
\hat{\bf B}f(z)=2^{-\frac{n}{4}}\int d\zeta e_*^{\frac{1}{2}\zeta\hat{G}\zeta-\zeta z-\frac{1}{4}z\hat{G}^{-1}z}f(z), \quad\quad z_k=G_{kj}p^j+\hat{\iota}\Theta_k.
\end{equation}
Therefore, one gets similar structures to the ones known for the fermionic Heisenberg group.
%
%
%
\section{Unitary representations of super-Heisenberg group}
In this section we want to briefly compare, presented above,  physical
description of super-Heisenberg group with the strict mathematical one. In the physical approach
it is, in many cases, enough to define supergroup as a supermatrix group and to work with formalism
where "representation" of the group is obtained with the use of the anti-commuting variables,
and the definition of the super-Hilbert space includes $Z_2$-graded algebra valued "scalar product".
In the strict approach the Lie supergroup is a group carrying
the supermanifold structure and the relevant representation theory is formulated in such a way
that notion of unitary representation can be used. The theory of unitary representations of
the nilpotent Lie supergroups is surprisingly new, it was formulated in 2006, comparing to the beginings
of the supersymmetry, which was invented in the Seventies of the last Century. This somehow gauges the scale of difficulties related with the supergroups.
The progress achieved in the last decade is related to the use of the simplification coming from identification of the super Lie group with the super Harish-Chandra pair.

Let us make short comparison of  differences in definitions of the main object used in the physical and strict approaches respectively.
%

\subsection{Super Lie group vs. super Harish-Chandra pair}
In the strict approach instead of the explicit supermatrix groups one considers the Lie supergroup: object which is a group and supermanifold of dimension $(n|m)$, so  locally it can be viewed as  $(R^n, C^{\infty}\otimes \bigwedge(\theta_1,\dots, \theta_m))$.
This object is not easy to handle but can be replaced by the super Harish-Chandra pair:  a pair $(G_0, \mathfrak{g})$, where $G_0$ is a Lie group with the lie algebra $\mathfrak{g}_0$ and $\mathfrak{g}=\mathfrak{g}_0\oplus\mathfrak{g}_1$ is a Lie superalgebra, moreover there are imposed  compatibility conditions:
\begin{itemize}
\item  $G$ acts on $\mathfrak{g}$ by $R$-linear automorphisms
\item
$
\sigma(G_0)|_{\mathfrak{g}_0}=Ad, \quad\quad d\sigma(X)Y=[X,Y]
$
\end{itemize}
To make  contact with our previous considerations, let us illustrate the super Harish-Chandra pair
notion on the example of  the super Heisenberg group \cite{kwok}
$SH_3=(H_3, \mathfrak{sh}(3|1))$
\begin{align}\nonumber
H_3 = \{
\left(
 \begin{array}{ccc|c}
 1 & x & t &0\\
 0 & 1 & y &0\\
 0 & 0 & 1 &0\\
 \hline
 0&0&0&1
 \end{array}
 \right)
 : x,y,t\in R
 \}
\end{align}
\begin{align}\nonumber
\mathfrak{sh}(3|1) = \{
\left(
\begin{array}{ccc|c}
0 & a & c &\alpha \\
0 & 0 & b & 0 \\
0 & 0 & 0 & 0 \\
\hline
0 & 0 & 0 & 0
\end{array}
\right)
: a,d,c,\alpha\in R
\}
\end{align}
Details of the unitary representation of such defined super-Heisenberg group can be found in Ref. \cite{kwok}.
%
\subsection{Unitary representations of super Lie groups}

There are various definitions of super-Hilbert space used in the literature. The definition of the super-Hilbert space relevant to study  unitary representations of the Lie supergroups has been observed in the context of the supercoherent state representations \cite{elgra_net}, and has been used in mathematical approach to supersymmetry \cite{deligne, salma,  kwok}. In such an approach the super-Hilbert space is defined as a
$Z_2$-graded complex Hilbert space $\mathcal{H}=\mathcal{H}_0\oplus \mathcal{H}_1 $
such that
$$
<v,\, w>=
\begin{cases}
0, &|v|=|w|+1\\
(v,w), &v,w\in\mathcal{H}_0\\
i(v,w), & v,w\in\mathcal{H}_1
\end{cases}
$$
where $(.,.)$ idenotes the scalar product in  $\mathcal{H}$ and $<.,.>$ is the
even super Hermitian form: $\overline{<v,w>}=(-1)^{|v||w|}<w,v>$.

On such  super-Hilbert spaces there are defined unitary representations of Lie supergroups\cite{cctv}. In particular, as it was shown by Salmasian in 2010 \cite{salma}, there is valid the generalization of the Stone-von Neumann to the case of the super-Heisenberg group i.e. the super Stone-von Neuman theorem \cite{salma}. Namely,
let $\chi:R\rightarrow C^{\times}$ be defined by $\chi(t)=e^{i\beta t}$, where $\beta>0$ then for
the supersymplectic form $\Omega$ we have  implications
\begin{itemize}
\item if  $\Omega|_{V_1\times V_1}$ is positive definite then exists a unique unitary representation of $SH$ with character $\chi$.
\item if $\Omega|_{V_1\times V_1}$ is not positive definite then $SH$ does not have any unitary representation with character $\chi$.
\end{itemize}

\section{Final Comments}
We have described in brief, how the idea of the representation of the super-Heisenberg group and the generalization of the harmonic analysis on it can be realized. One approach, uses ideas which are present in the formulation of the supersymmetric field theory. Such physical approach yields the supermodule over superalgebra description of the representations. The strict mathematical approach allows to keep the notion of unitary representation. It turned out to be much more difficult to obtain and has been formulated only recently.
%

\end{document}